\documentclass[letter]{aa} 

\usepackage{textcomp}
\usepackage{graphicx} %
\usepackage{amsmath} %
\usepackage{amssymb} %
\usepackage{bm} %
\usepackage{upgreek} %
\usepackage{IEEEtrantools} %
\usepackage{multirow}
\usepackage[colorlinks=true,allcolors=blue,urlcolor=blue]{hyperref}
\usepackage{txfonts}
\usepackage[normalem]{ulem} 
\usepackage{multirow}

\newcommand{\sect}[1]{\text{Sect.~\ref{#1}}}
\newcommand{\fig}[1]{\text{Fig.~\ref{#1}}}

\newcommand{\multitd}{\textsc{multi3d}}
\newcommand{\balder}{\textsc{balder}}

\newcommand{\marcs}{\textsc{marcs}}

\newcommand{\stagger}{\textsc{stagger}}

\newcommand{\teff}{T_{\mathrm{eff}}}
\newcommand{\lgg}{\log{g}}

\newcommand{\feh}{\mathrm{\left[Fe/H\right]}}
\newcommand{\xfe}[1]{\mathrm{\left[#1/Fe\right]}}

\newcommand{\abrat}[2]{\mathrm{\left[#1/#2\right]}}
\newcommand{\lgeps}[1]{\log{\epsilon_{\mathrm{#1}}}}

\newcommand{\dex}{\mathrm{dex}}

\newcommand{\hbeta}{\mathrm{H\upbeta}}

\newcommand{\nm}{\mathrm{nm}}
\begin{document} 

\title{Carbon and oxygen in metal-poor halo stars\thanks{Tables 1--4 
are available in electronic form at the CDS
via anonymous ftp to 
\url{cdsarc.u-strasbg.fr (130.79.128.5)} or 
via~\url{http://cdsarc. u-strasbg.fr/viz-bin/qcat?J/A+A/622/L4}.}}
\author{A.~M.~Amarsi\inst{1}
\and
P.~E.~Nissen\inst{2}
\and
M.~Asplund\inst{3,4}
\and
K.~Lind\inst{1,5}
\and
P.~S.~Barklem\inst{6}}
\institute{Max-Planck-Institut f\"ur Astronomy, K\"onigstuhl 17, 
D-69117 Heidelberg, Germany\\
\email{amarsi@mpia.de}
\and
Stellar Astrophysics Centre, Department of Physics and Astronomy, Aarhus
University, Ny Munkegade 120, DK-8000 Aarhus C, Denmark
\and
Research School of Astronomy and Astrophysics, Australian National
University, Canberra, ACT 2611, Australia
\and
ARC Centre of Excellence for 
All Sky Astrophysics in 3 Dimensions (ASTRO 3D), Australia
\and
Observational Astrophysics, Department of Physics and Astronomy, 
Uppsala University, Box 516, SE-751 20 Uppsala, Sweden
\and
Theoretical Astrophysics, Department of Physics and Astronomy, 
Uppsala University, Box 516, SE-751 20 Uppsala, Sweden}

\abstract{Carbon and oxygen are key tracers
of the Galactic chemical evolution;  in particular,
a reported upturn in $\abrat{C}{O}$~towards 
decreasing $\abrat{O}{H}$~in metal-poor halo stars
could be a signature of 
nucleosynthesis by massive Population III stars.
We reanalyse carbon, oxygen,
and iron abundances in 39 metal-poor turn-off stars.
For the first time,
we take into account 3D hydrodynamic effects
together with departures from local thermodynamic equilibrium (LTE)
when determining both the stellar parameters and the elemental abundances,
by deriving effective temperatures from 3D non-LTE 
$\hbeta$~profiles, surface gravities from Gaia parallaxes,
iron abundances from 3D LTE \ion{Fe}{II}~equivalent widths,
and carbon and oxygen
abundances from 3D non-LTE \ion{C}{I}~and \ion{O}{I}~equivalent widths.
We find that $\abrat{C}{Fe}$~stays flat with $\feh$,
whereas $\abrat{O}{Fe}$~increases linearly up to $0.75\,\dex$~with decreasing
$\feh$~down to $-3.0\,\dex$.
Therefore $\abrat{C}{O}$~monotonically decreases
towards decreasing $\abrat{O}{H}$,
in contrast to previous findings, mainly 
because the non-LTE effects for \ion{O}{I}
at low $\feh$~are weaker with our improved calculations.}

\maketitle

\section{Introduction}
\label{introduction}

Owing to their different formation sites
with different production timescales,
the abundance ratios of carbon, oxygen, and iron are key tracers 
of the chemical evolution of our Galaxy \citep{1979ApJ...229.1046T}.
Carbon in the cosmos comes from low- and intermediate-mass 
stars, and through core- and shell-burning in massive stars,
and it may be released through core-collapse supernovae
as well as through metallicity-dependent stellar winds;
oxygen is mainly produced in hydrostatic burning in massive stars
and is released through core-collapse supernova;
and iron is produced in both core-collapse and type Ia
supernova \citep[e.g.][]{2003MNRAS.339...63C,
2009A&amp;A...505..605C,2009ApJ...707.1466K,2014PASA...31...30K}.

In metal-poor halo stars with $\feh\lesssim-1.5$, 
the $\abrat{C}{O}$\footnote{$\abrat{A}{B}\equiv
\log_{10}\frac{N_{\mathrm{A}}}{N_{\mathrm{B}}}-
\left(\log_{10}\frac{N_{\mathrm{A}}}{N_{\mathrm{B}}}\right)_{\odot}$}~against 
$\abrat{O}{H}$~trend
has been studied in detail by
\citet{2004A&amp;A...414..931A} and \citet{2009A&amp;A...500.1143F}.
They found that $\abrat{C}{O}$~decreases
with decreasing $\abrat{O}{H}$, down to around $\abrat{O}{H}\approx-1.0$.
This is qualitatively consistent with what was found more recently
by \citet{2014A&amp;A...568A..25N} in the less
metal-poor halo as well as in
thick-disk stars with $\feh\gtrsim-1.5$;
there, $\abrat{C}{O}$~decreases 
from $0.0\,\dex$~at $\abrat{O}{H}\approx0.2$,
down to $-0.45\,\dex$~at $\abrat{O}{H}\approx-0.4$.
This trend could signal
the presence of a metallicity-dependent carbon yield from 
the winds of massive stars or an increasing contribution
of carbon from low- and intermediate-mass stars with cosmic time.

At lower $\abrat{O}{H}$, 
\citet{2004A&amp;A...414..931A} and \citet{2009A&amp;A...500.1143F}
found evidence for an overturn in the trend: $\abrat{C}{O}$~increases with further decreasing 
$\abrat{O}{H}$, reaching $\abrat{C}{O}\approx0.0$~at
$\abrat{O}{H}\approx-3.0$.
As discussed in these studies,
one interpretation of the change in behaviour in $\abrat{C}{O}$~at 
low $\abrat{O}{H}$~is that it is a signature of
nucleosynthesis by massive Population III stars because the yields of these massive, metal-free first stars 
are relatively rich in carbon
\citep[e.g.][]{2014ApJ...792L..32I}.
Alternative interpretations are also possible,
for example, that it could signal fast stellar rotation in metal-poor
Population II stars \citep[][]{2006A&A...447..623M,2006A&A...449L..27C}.

However, there are a number of ways that the 
stellar parameter and elemental abundance
determinations of these earlier works can be improved.
The effective temperatures, and carbon, oxygen, and iron abundances
in these older analyses were based on
1D hydrostatic model atmospheres.
Furthermore, while departures from local thermodynamic equilibrium
(LTE) were later taken into account for \ion{C}{I} and \ion{O}{I},
these calculations were based on older atomic data
antecedent to modern descriptions
of the inelastic collisions with neutral hydrogen
\citep[e.g.][]{2016A&amp;ARv..24....9B}.

Here we revisit the $\abrat{C}{O}$~against $\abrat{O}{H}$~trend
in the metal-poor halo.
We present for the first time an abundance analysis
that is based on both 3D non-LTE stellar parameters
and 3D non-LTE elemental abundances that are based on
the best atomic data currently available.

\section{Method}
\label{method}

\subsection{Overview}
\label{methodoverview}

The sample consists of 39 of the 40 stars
that have been observed with the VLT/UVES echelle spectrograph
by \citet{2007A&amp;A...469..319N};
G 066-030 was not included here
because the stellar parameters we derived 
($\teff=6596\,\mathrm{K}$, $\lgg=4.70$, $\feh=-1.24$)
suggest that the star is a blue straggler
\citep[as also suggested by][]{2017A&amp;A...608A..46R}.
The sample of \citet{2009A&amp;A...500.1143F} contains
additional stars that are not considered here:
G 041-041, G 084-029, and LP 831-070 (for the last of which
only an upper limit on the oxygen abundance could be obtained).
We only consider the stars from
\citet{2007A&amp;A...469..319N} in order to base the analysis
on a homogeneous set of $\hbeta$~observations.

The stellar parameters (effective temperatures, surface gravities,
and iron abundances) were determined prior to performing
the abundance analysis of carbon and oxygen.
They were determined by the separate methods described below,
and iterated until consistency was achieved.
Following \citet{2015MNRAS.454L..11A},
line formation calculations were performed on the
\stagger-grid of 3D hydrodynamic model atmospheres
\citep{2013A&amp;A...557A..26M},
and for comparison also on the standard grid of
\marcs~1D hydrostatic model atmospheres
\citep{2008A&amp;A...486..951G}.
These adopt the solar chemical compositions
from \citet{2009ARA&amp;A..47..481A}
and \citet{2007coma.book..105G},
respectively, scaled by $\feh$~such that 
$\xfe{\upalpha}=0.4$~for $\upalpha$-elements and 
$\xfe{X}=0.0$~for other elements.
These compositions were
also assumed in the line formation calculations (except 
for carbon and oxygen, in \sect{methodabundances}).
This is a reasonable assumption, given that
hydrogen is the dominant electron donor across the atmospheres
of turn-off stars with $\feh\lesssim-1.5$;
We caution, however, that peculiar 
$\xfe{Mg}$~and $\xfe{Si}$~abundances may add to the scatter
in the more metal-rich part of our sample.

All line formation calculations were performed 
using the 3D non-LTE radiative transfer code
\balder~\citep{2018A&A...615A.139A}, our custom version
of \multitd~\citep{2009ASPC..415...87L_short}.
When using the same stellar parameters and equivalent widths,
our 1D LTE carbon and oxygen abundances agree with those
of \citet{2009A&amp;A...500.1143F} to $0.015\,\dex$, on average;
the differences tend to go in the same direction and leave
$\abrat{C}{O}$~even less affected.
We will present our grids of synthetic 3D non-LTE equivalent widths 
in a future paper. We list the stellar parameters and abundances
in the online Tables 1--4.

\subsection{Effective temperature}
\label{methodteff}

The effective temperatures were determined by performing
profile fits of $\hbeta$, using 
the spectra of \citet{2007A&amp;A...469..319N}
and the 3D non-LTE grid of \citet{2018A&A...615A.139A}.
The continuum placement and effective temperature were
fit simultaneously by $\chi^{2}$-minimisation, 
given the surface gravity and iron abundance.
The fitting windows included the region within
$2.4\,\nm$~from line centre, excluding
the region within $0.2\,\nm$~from line centre
because the line core forms in the chromosphere and is only poorly modelled by this 3D non-LTE grid.

The new 3D non-LTE effective temperatures tend to be lower than
the 1D LTE effective temperatures of \citet{2007A&amp;A...469..319N}, as expected 
from the discussion of $\hbeta$~in \citet{2018A&A...615A.139A}.
In general, the difference is larger for the cooler stars
(up to $80\,\mathrm{K}$, at $\teff\approx5800\,\mathrm{K}$)
than for the hotter stars
(around $10\,\mathrm{K}$~at $\teff\approx6400\,\mathrm{K}$).

\subsection{Surface gravity}
\label{methodlgg}

Stellar masses, effective temperatures, and absolute bolometric magnitudes were
used to determine surface gravities as described in 
\citet[][Sect.~3.2]{2007A&amp;A...469..319N},
with two improvements.  First, and most important, absolute visual
magnitudes were determined based on Gaia DR2 parallaxes
\citep{2018A&A...616A...1G}.
Two stars have no Gaia parallax available: HD 84937, for which we
adopted the HST parallax \citep{2014ApJ...792..110V},
and CD-35 14849, for which
the photometric absolute visual magnitude based on 
Str{\"o}mgren photometry was used. 
Second, a newer calibration of the bolometric correction as a function of
$V-K$~\citep{2010A&amp;A...512A..54C} was adopted.

Thanks to the high precision of Gaia parallaxes, the new surface gravities are
estimated to have RMS errors of about $\pm 0.05\,\dex$.  In comparison,
the gravities in \citet[][]{2007A&amp;A...469..319N}
were estimated to have errors of $\pm 0.15\,\dex,$~with the
main uncertainty arising from errors in the Hipparcos
parallaxes and the estimates of absolute magnitudes
from Str{\"o}mgren photometry.

\subsection{Iron abundance}
\label{methodfeh}

The iron abundances were determined from 
equivalent widths of \ion{Fe}{II}
lines measured in VLT/UVES echelle spectra
\citep{2002A&amp;A...390..235N,2004A&amp;A...415..993N,
2007A&amp;A...469..319N}.
On average, 16 different optical
\ion{Fe}{II}~lines were available for a given
star (only 3 were available for G 064-012 and G 064-037).
As the non-LTE effects on \ion{Fe}{II}~lines 
are expected to be small \citep[e.g.][]{2016MNRAS.463.1518A},
the literature equivalent widths were compared to our grid of 
3D LTE \ion{Fe}{II} equivalent widths.
This grid was constructed based on
the line parameters from \citet{2009A&amp;A...497..611M}.
We adopt for the solar iron abundance 
$\lgeps{Fe}=7.50$~\citep{2009ARA&amp;A..47..481A},
consistent with the 3D model atmospheres.

The new 3D LTE iron abundances tend to be slightly larger than the 
old 1D LTE abundances, as expected from previous 3D LTE studies
\citep[e.g.][]{2016MNRAS.463.1518A}.
The differences tend to be larger for the cooler stars
(up to $+0.14\,\dex$) than 
for the hotter stars ($+0.01\,\dex$).

\subsection{Carbon and oxygen abundances}
\label{methodabundances}

The carbon and oxygen abundances were determined 
from equivalent widths of \ion{C}{I} and \ion{O}{I}
lines measured in VLT/UVES echelle spectra
given in \citet{2002A&amp;A...390..235N},
\citet{2009A&amp;A...500.1143F},
and \citet{2004A&amp;A...414..931A},
in order of preference.
The spectra have $R\approx60000$~and 
signal-to-noise ratios of $200$~to $300$, and the equivalent
widths have precisions of around $0.2\,\mathrm{pm}$;
furthermore, the continuum is well defined, as is shown in 
Fig.~2 of \citet[][]{2004A&amp;A...414..931A}.
The lines are all of high-excitation potential and in the 
optical/infra-red
(\ion{C}{I} $906.1\,\nm$, $906.2\,\nm$,
$907.8\,\nm$, $908.9\,\nm$, $909.5\,\nm$,
$911.2\,\nm$, $940.6\,\nm$; \ion{O}{I} $777.2\,\nm$,
$777.4\,\nm$, and $777.5\,\nm$).
Their sensitivities to the stellar parameters
therefore cancel out in the ratio $\abrat{C}{O}$, at least to first order.

The literature equivalent widths were compared to our grids of 
1D LTE, 1D non-LTE,
3D LTE, and 3D non-LTE \ion{C}{I} and \ion{O}{I} equivalent widths.
The model \ion{C}{I} and \ion{O}{I} atoms and 
3D non-LTE procedure were recently presented by
Amarsi et al. (in prep.) and \citet{2018A&A...616A..89A};
the oscillator strengths of the diagnostic lines are from 
\citet{1991JPhB...24.3943H,1993A&AS...99..179H},
via the NIST Atomic Spectra Database \citep{NIST_ASD}.
We adopt for the solar abundances 
$\lgeps{C}=8.43$ and $\lgeps{O}=8.69$~\citep{2009ARA&amp;A..47..481A}.

\begin{figure*}
    \begin{center}
        \includegraphics[scale=0.31]{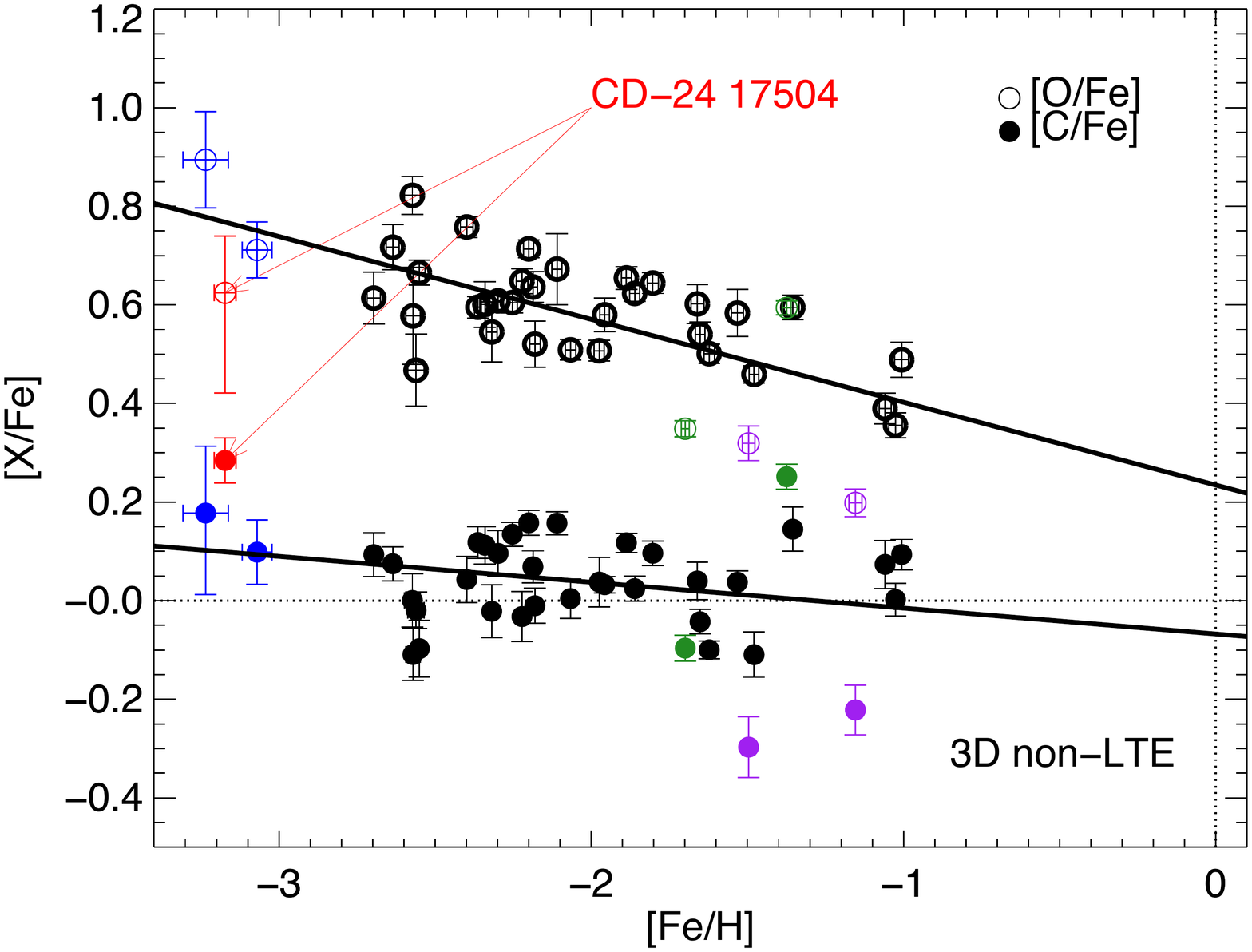}
        \includegraphics[scale=0.31]{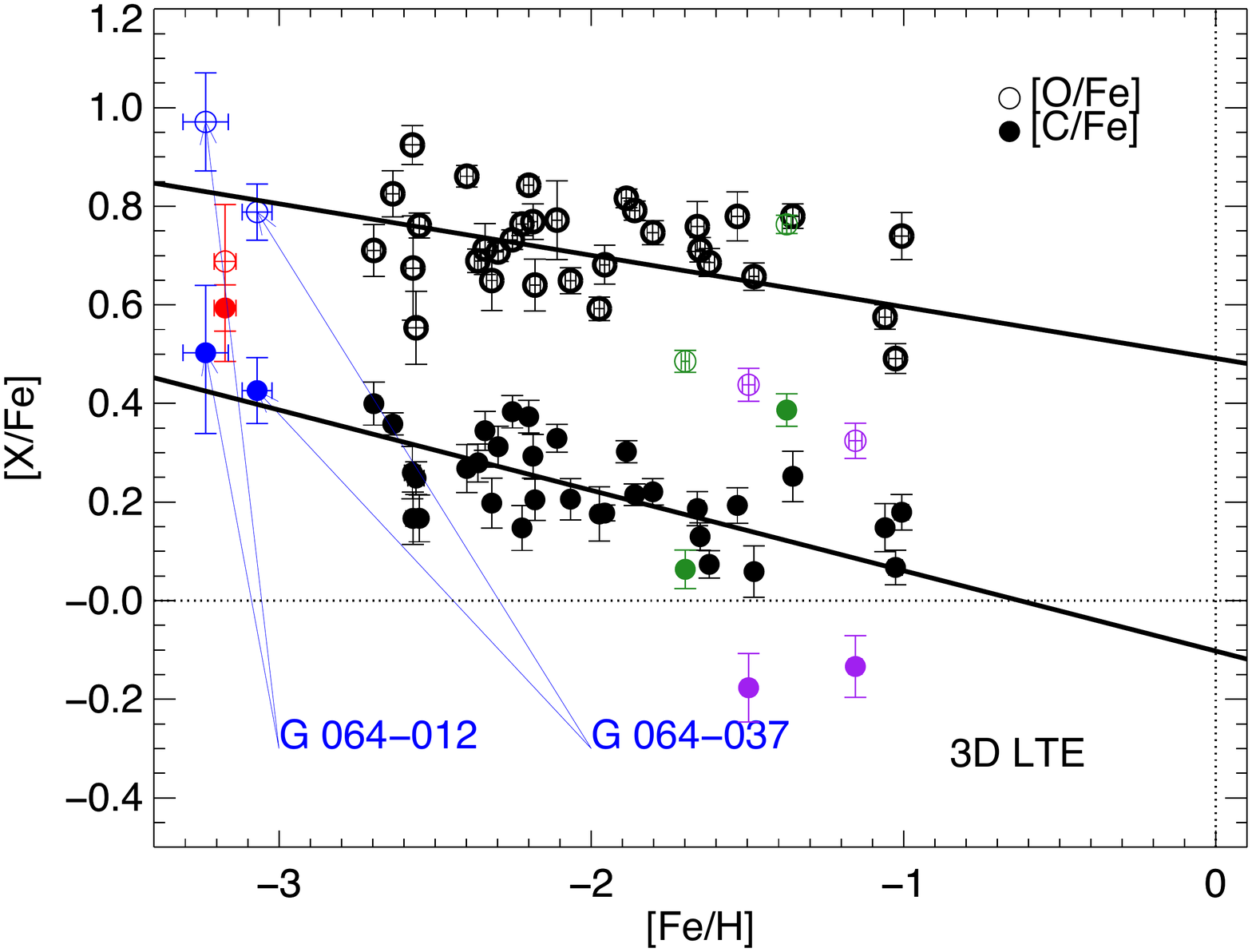}
        \includegraphics[scale=0.31]{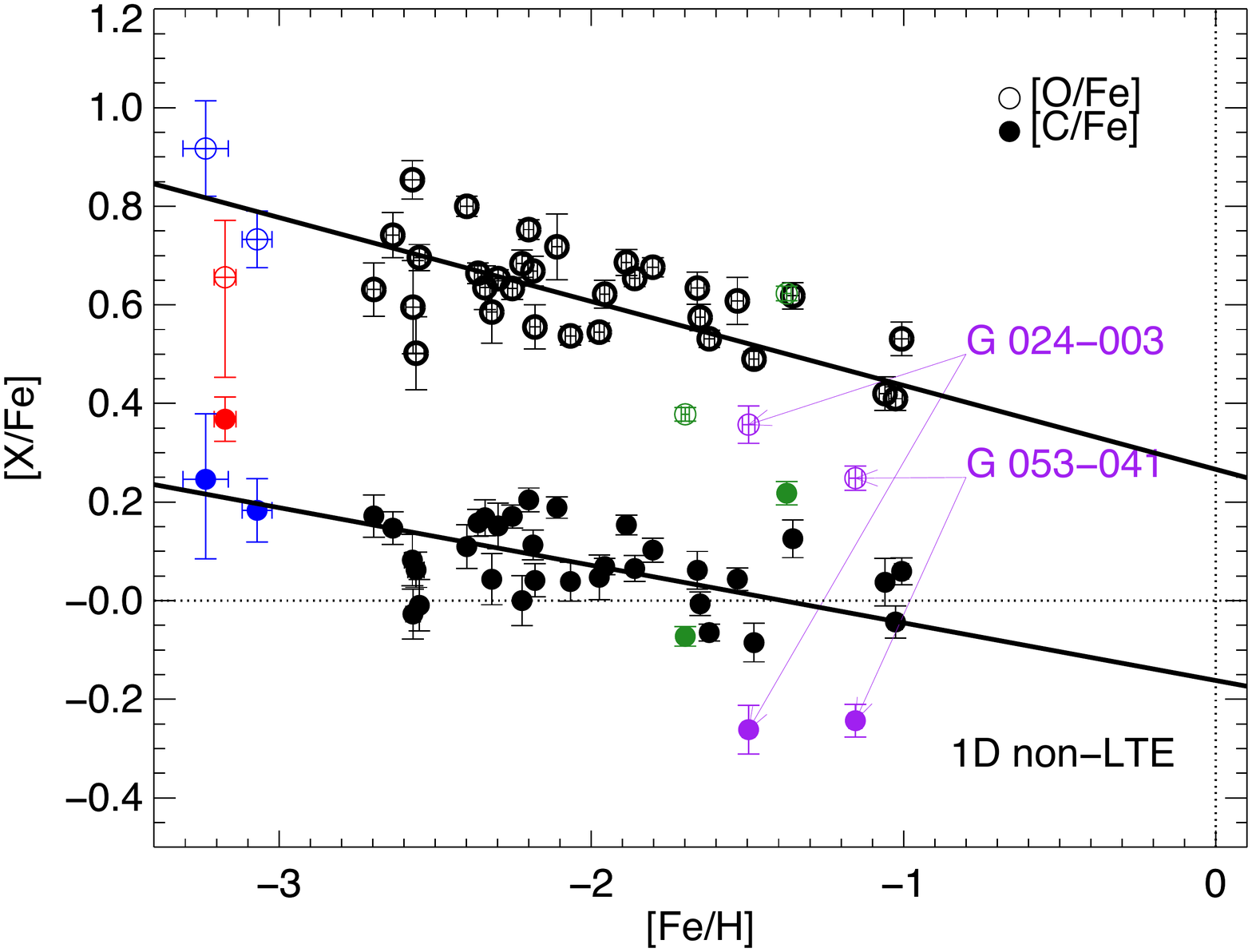}
        \includegraphics[scale=0.31]{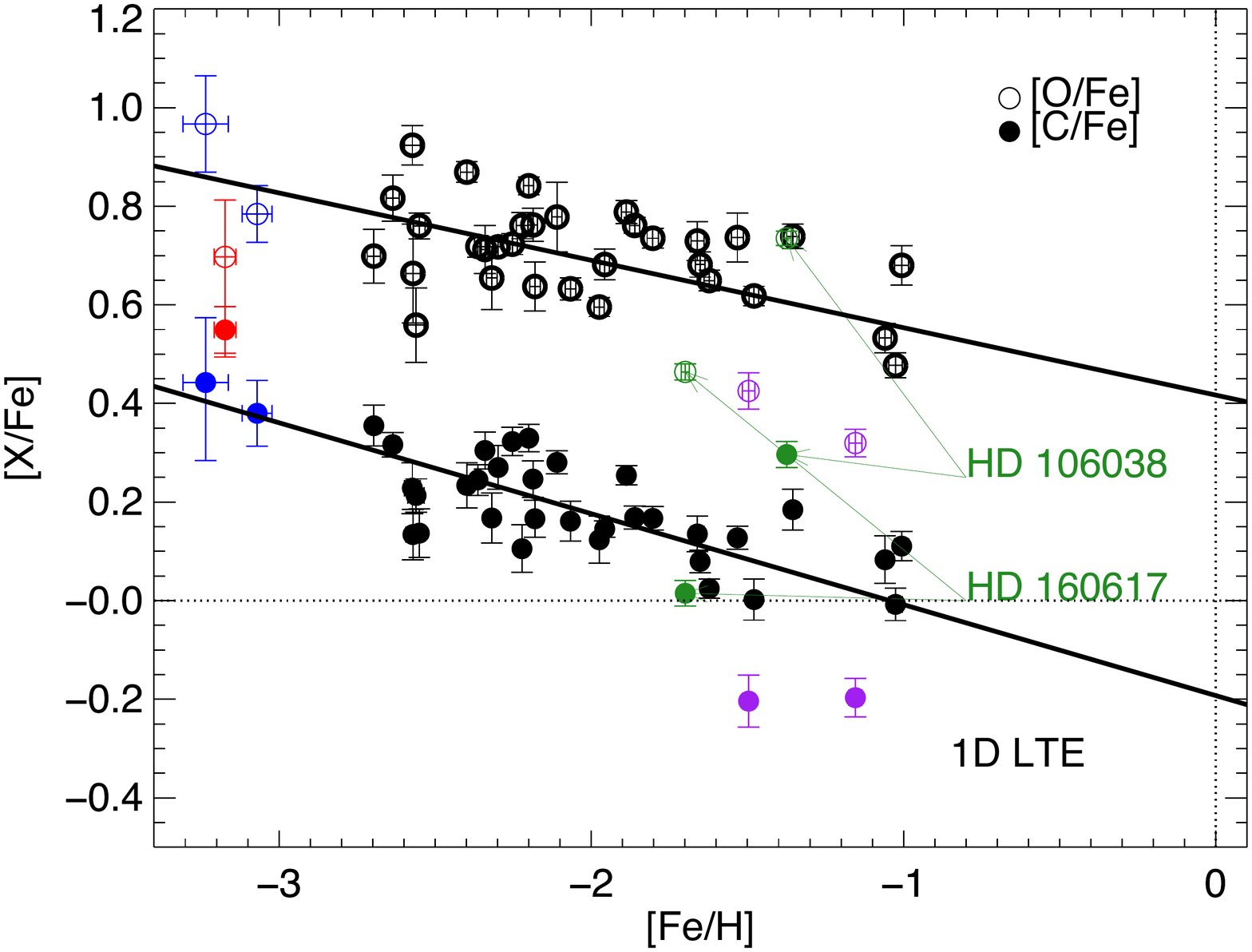}
        \caption{$\abrat{C}{Fe}$~and $\abrat{O}{Fe}$~against 
        $\abrat{Fe}{H}$~for
        different line formation calculations of $\ion{C}{I}$~and
        $\ion{O}{I}$; $\teff$, $\lgg$, and $\feh$~were fixed to their
        3D non-LTE values in all four panels.
        Weighted linear fits are also shown.}
        \label{fig:cfe_ofe_feh}
    \end{center}
\end{figure*}

\begin{figure*}
    \begin{center}
        \includegraphics[scale=0.31]{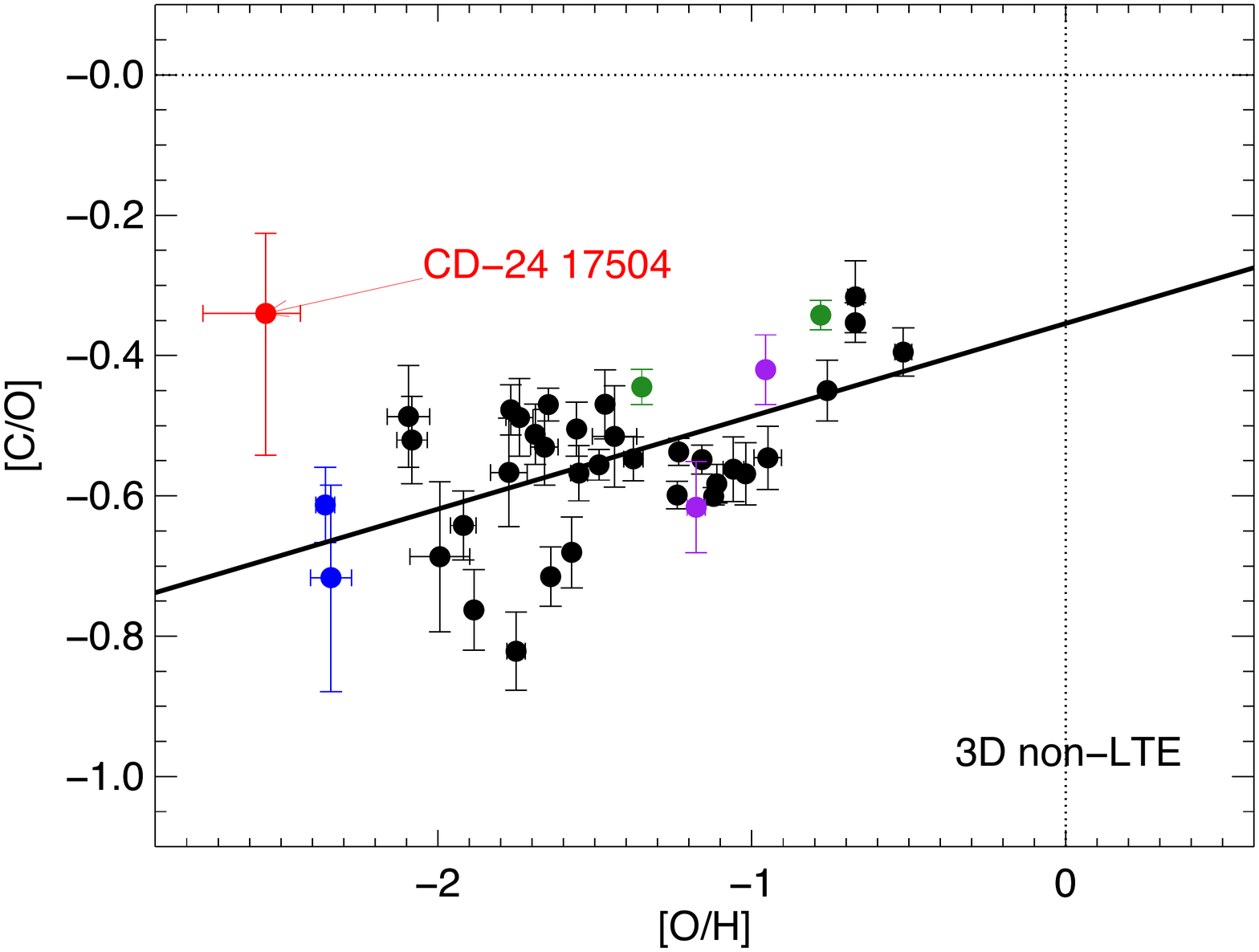}
        \includegraphics[scale=0.31]{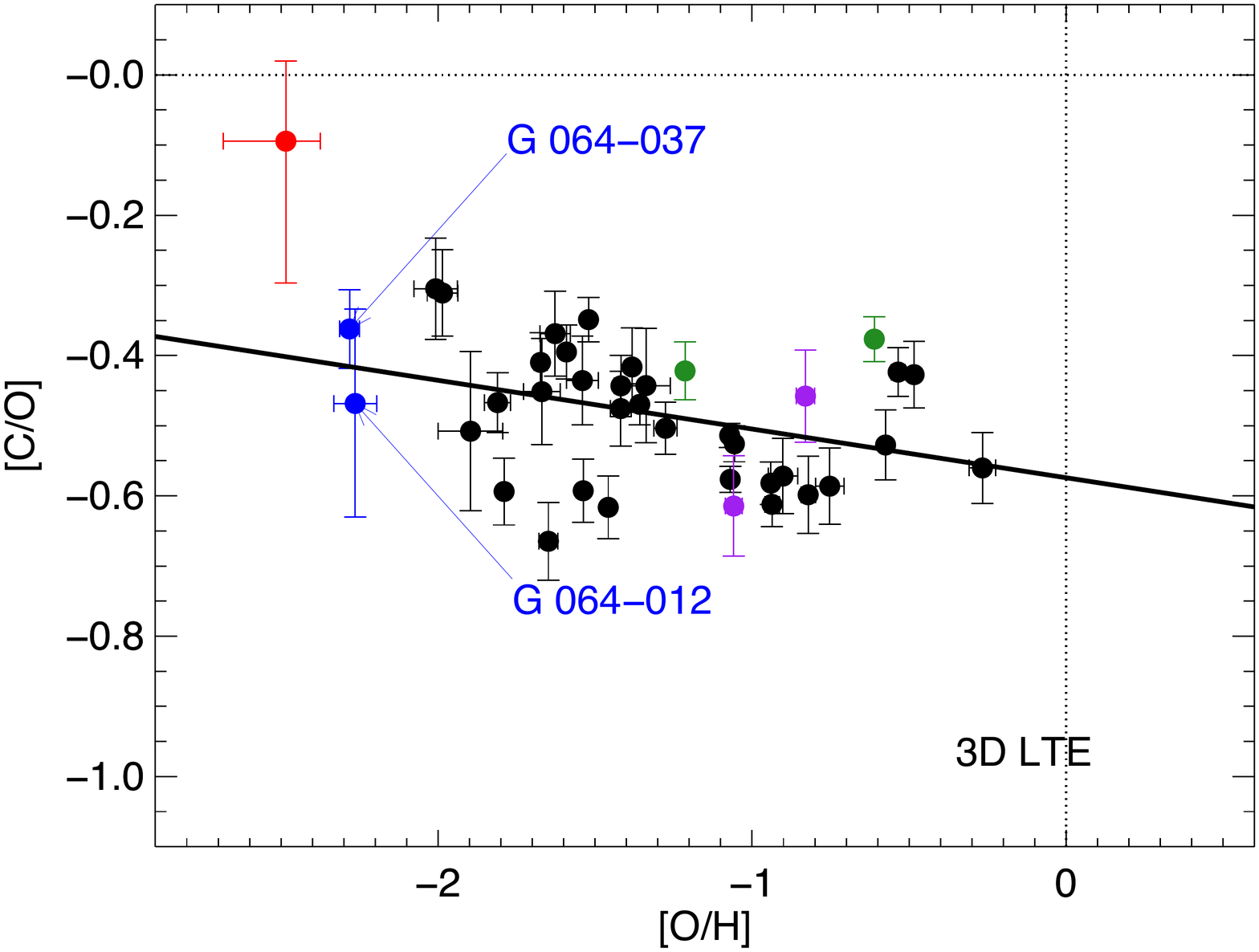}
        \includegraphics[scale=0.31]{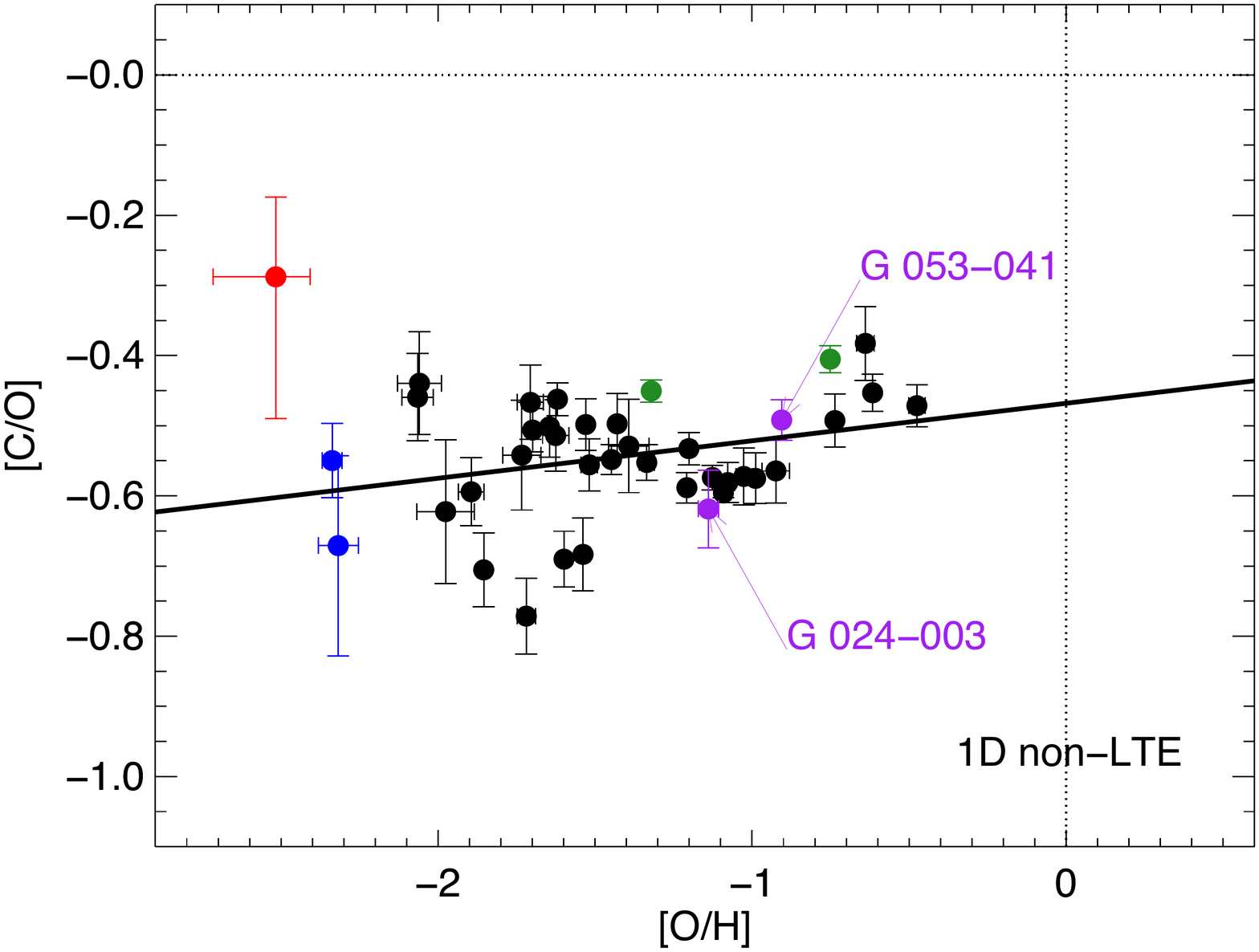}
        \includegraphics[scale=0.31]{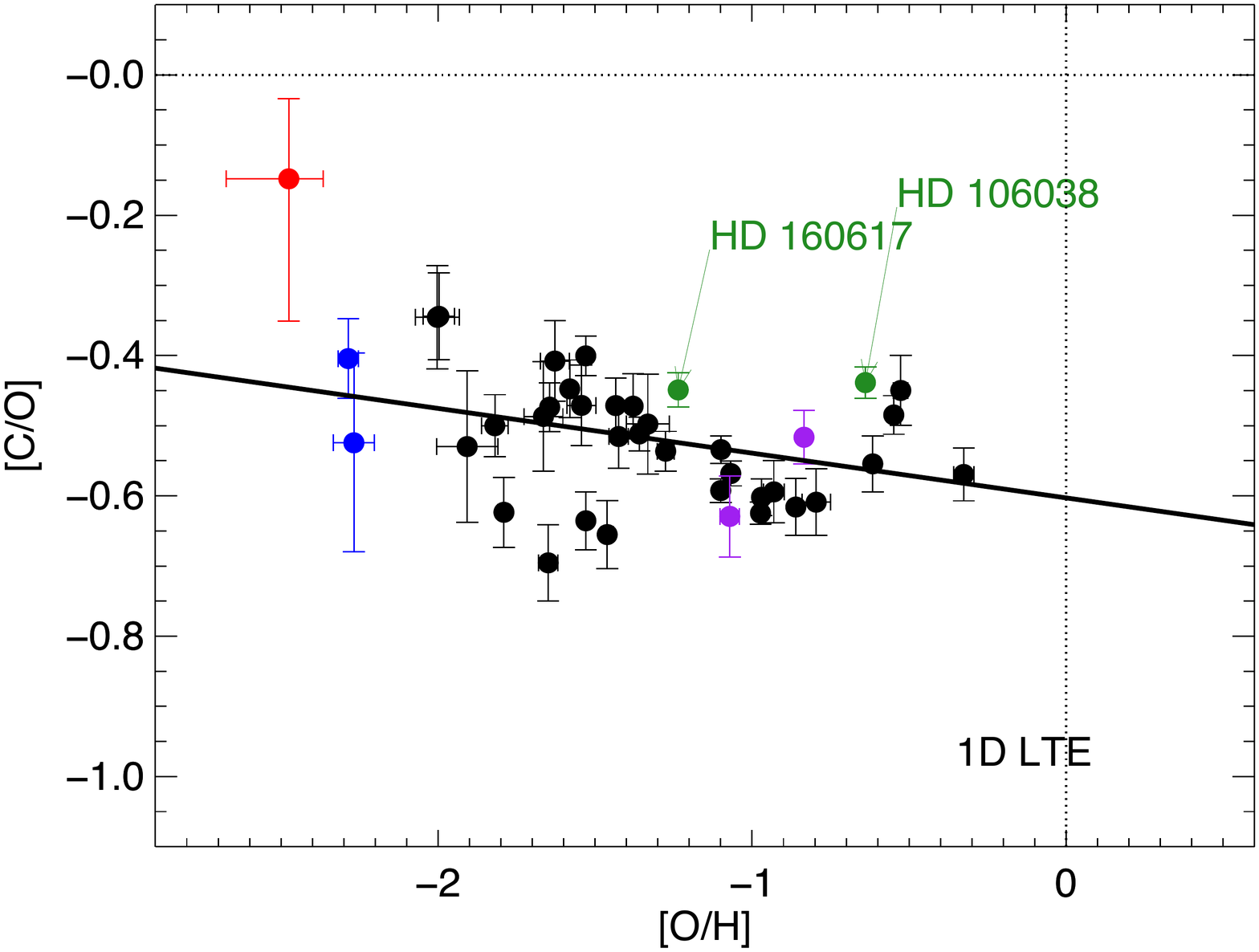}
        \caption{$\abrat{C}{O}$~against $\abrat{O}{H}$~for
        different line formation calculations of $\ion{C}{I}$~and
        $\ion{O}{I}$; $\teff$, $\lgg$, and $\feh$~were fixed to their
        3D non-LTE values in all four panels. 
        Weighted linear fits are also shown.}
        \label{fig:co_oh}
    \end{center}
\end{figure*}

\section{Results}
\label{results}

We illustrate the results in \fig{fig:cfe_ofe_feh}
and \fig{fig:co_oh}.
The error bars indicate the statistical error in the mean
based on the line-to-line scatter.
The carbon abundance of G 064-012 was determined from a single line,
which means that the error bars
here reflect an uncertainty of $0.23\,\mathrm{pm}$~on
the equivalent width of the \ion{C}{I} $909.5\,\nm$ line
($0.87\,\mathrm{pm}$), based on a scatter of $0.32\,\mathrm{pm}$~in
the correlation between the measured 
equivalent widths of the \ion{C}{I} $909.5\,\nm$ and 
$940.6\,\nm$~lines in the rest of the sample.
The oxygen abundance of CD-24 17504 was also determined from a single line,
therefore the error bars here reflect uncertainties of $+0.11$/$-0.20\,\dex$~in
$\abrat{O}{H}$, based on the spectrum analysis of
\citet[][Fig.~4]{2009A&amp;A...500.1143F}.

\fig{fig:cfe_ofe_feh} shows that the 
3D non-LTE analysis indicates that on average,
$\abrat{C}{Fe}$~remains flat with $\feh$, 
all the way down to $\abrat{Fe}{H}\approx-3.0$,
with a mean value of around $0.1\,\dex$.
On the other hand, $\abrat{O}{Fe}$~monotonically increases
with decreasing $\feh$, with
$\abrat{O}{Fe}\approx0.4$~at $\feh\approx-1.0$,
increasing to $\abrat{O}{Fe}\approx0.75$~at $\feh\approx-3.0$.

\fig{fig:co_oh} shows that 
the 1D LTE and 3D LTE $\abrat{C}{O}$~against
$\abrat{O}{H}$~trends are very similar:
$\abrat{C}{O}$~slightly increases towards lower $\abrat{O}{H}$.
In contrast, in non-LTE, $\abrat{C}{O}$~monotonically
decreases with decreasing $\abrat{O}{H}$;
the gradient is clearly steeper for the 3D non-LTE results
than for the 1D non-LTE results.

By comparing the different panels,
\fig{fig:cfe_ofe_feh} also shows 
that the 3D effects and the non-LTE effects go in the same direction,
acting to reduce 3D non-LTE abundances
compared to 1D LTE, 3D LTE, and 1D non-LTE;
this is consistent with our earlier findings for oxygen
\citep[][]{2016MNRAS.455.3735A}.  
The 3D non-LTE effects in \ion{C}{I}~are more severe
towards lower $\feh$, whereas the effects in \ion{O}{I}~are more
severe towards higher $\feh$.
This drives the steep negative gradient
in the 3D non-LTE $\abrat{C}{O}$~trend in
\fig{fig:co_oh} compared to the 1D LTE result.

\section{Discussion}
\label{discussion}

Based on the 3D non-LTE results, we
find no evidence for an upturn in
$\abrat{C}{O}$~at low $\abrat{O}{H}$;
rather, $\abrat{C}{O}$~decreases monotonically
between $2.6<\abrat{O}{H}<-0.5$. This result
is in contrast to the upturn found in the earlier studies
of \citet{2004A&amp;A...414..931A} and \citet{2009A&amp;A...500.1143F}.
The main reason for the difference with \citet{2009A&amp;A...500.1143F}
is that our models give
far weaker departures from LTE in \ion{O}{I}~in the metal-poor
regime for the reasons discussed in 
\citet[][Sect.~4.2]{2016MNRAS.455.3735A}.
Thus, in so far as an overturn in the relation would
signal Population III nucleosynthesis, we do 
not find any Population III signature
in the $\abrat{C}{O}$~against $\abrat{O}{H}$~plane
in this sample of metal-poor halo stars.

It is likely that the mean trends are influenced by an 
intrinsic cosmic scatter.  
For instance, HD 106038 and HD 160617 both
lie above the mean $\abrat{C}{O}$~trends because of somewhat high $\abrat{C}{Fe}$ and
low $\abrat{O}{Fe}$, respectively.
The former star may have been influenced
by a hypernova event \citep{2008MNRAS.385L..93S},
while the latter star is boron poor and nitrogen rich
\citep{2014ApJ...784..158R}, whose abundances are 
probably affected by surface mixing \citep{1997ARA&amp;A..35..557P}.
In addition, although stars G 053-041 and G 024-003
lie near the mean $\abrat{C}{O}$~trends, their $\abrat{C}{Fe}$~and $\abrat{O}{Fe}$~are  both lower by roughly
$0.2\,\dex$~than the mean trends.
The former star was previously found to be 
sodium enhanced \citep{2010A&amp;A...511L..10N}
and was likely born in a globular cluster
\citep{2012ApJ...757..164R}; we speculate
that the latter star may have similar origins.

The 1D LTE and
3D non-LTE results both imply that none of the stars in this sample qualify 
as carbon-enhanced metal-poor stars (CEMP; $\xfe{C}>0.7$).
The most carbon-enhanced star in the sample is
CD-24 17504: our 3D non-LTE result is 
$\xfe{C}=0.28$.
This is significantly lower than
$\xfe{C}=1.1$~reported by \citet{2015ApJ...808...53J}, which was based on a 1D analysis of CH lines in the G band.
Similarly, for G 064-012 and G 064-037, we infer
$\xfe{C}=0.18$~and $\xfe{C}=0.10$, respectively,
significantly lower than the values of
$\xfe{C}=1.07$~and $\xfe{C}=1.12$~reported by
\citet{2016ApJ...829L..24P}, which were
based on a 1D analysis of CH lines in the G band.
Molecular features are highly susceptible
to 3D effects, with abundance corrections
becoming more negative towards lower
$\feh$~that are of the right order
of magnitude ($0.7$~to $1.1\,\dex$) 
to bring their results into agreement
with ours \citep[e.g.][]{2006ApJ...644L.121C,
2017A&amp;A...598L..10G}.

Lastly, based on the \ion{O}{I}~$777\,\nm$~triplet,
\ion{Fe}{II}~lines, and 3D non-LTE stellar parameters,
we find that the $\xfe{O}$~against $\feh$~trend
does not plateau, but shows a clear monotonic decrease
with increasing $\feh$. 
This brings the infra-red triplet into agreement 
with measurements of UV OH lines in metal-poor red giant stars
\citep[][]{2015A&amp;A...576A.128D,2018MNRAS.475.3369C}.
However, it is difficult to reconcile with
measurements of the [\ion{O}{I}]~$630\,\nm$~line
in metal-poor stars \citep[][]{2015MNRAS.454L..11A}.
We will revisit this oxygen problem in metal-poor stars in a future work.


\begin{acknowledgements}
We thank the referee for valuable feedback on the manuscript.
AMA and KL acknowledge funds from the Alexander von Humboldt Foundation in
the framework of the Sofja Kovalevskaja Award endowed by the Federal Ministry of
Education and Research, and KL also
acknowledges funds from the Swedish Research Council 
(grant 2015-004153) and Marie Sk{\l}odowska Curie Actions 
(cofund project INCA 600398).
Funding for the Stellar Astrophysics Centre is provided by The Danish
National Research Foundation (grant DNRF106).
MA gratefully acknowledges funding from
the Australian Research Council (grants FL110100012 and DP150100250).  
Parts of this research were conducted by the 
Australian Research Council Centre of Excellence for 
All Sky Astrophysics in 3 Dimensions (ASTRO 3D),
through project number CE170100013.
PSB acknowledges financial support from the Swedish
Research Council and the project grant ``The New Milky Way'' from the Knut and
Alice Wallenberg Foundation.  
This work was based on observations collected
at the European Southern Observatory under ESO programmes 67.D-0106 and 
73.D-0024.
This work has made use of data from the European Space Agency (ESA) mission
{\it Gaia} (\url{https://www.cosmos.esa.int/gaia}), processed by the {\it Gaia}
Data Processing and Analysis Consortium (DPAC,
\url{https://www.cosmos.esa.int/web/gaia/dpac/consortium}). Funding for the DPAC
has been provided by national institutions, in particular the institutions
participating in the {\it Gaia} Multilateral Agreement.
This work was supported by computational resources provided by the Australian
Government through the National Computational Infrastructure (NCI)
under the National Computational Merit Allocation Scheme.
\end{acknowledgements}


\bibliographystyle{aa} 
\bibliography{/Users/ama51/Documents/work/papers/allpapers/bibl.bib}

\begin{thebibliography}{44}
\expandafter\ifx\csname natexlab\endcsname\relax\def\natexlab#1{#1}\fi

\bibitem[{{Akerman} {et~al.}(2004){Akerman}, {Carigi}, {Nissen}, {Pettini}, \&
  {Asplund}}]{2004A&amp;A...414..931A}
{Akerman}, C.~J., {Carigi}, L., {Nissen}, P.~E., {Pettini}, M., \& {Asplund},
  M. 2004, \aap, 414, 931

\bibitem[{{Amarsi} {et~al.}(2015){Amarsi}, {Asplund}, {Collet}, \&
  {Leenaarts}}]{2015MNRAS.454L..11A}
{Amarsi}, A.~M., {Asplund}, M., {Collet}, R., \& {Leenaarts}, J. 2015, \mnras,
  454, L11

\bibitem[{{Amarsi} {et~al.}(2016{\natexlab{a}}){Amarsi}, {Asplund}, {Collet},
  \& {Leenaarts}}]{2016MNRAS.455.3735A}
{Amarsi}, A.~M., {Asplund}, M., {Collet}, R., \& {Leenaarts}, J.
  2016{\natexlab{a}}, \mnras, 455, 3735

\bibitem[{{Amarsi} {et~al.}(2018{\natexlab{a}}){Amarsi}, {Barklem}, {Asplund},
  {Collet}, \& {Zatsarinny}}]{2018A&A...616A..89A}
{Amarsi}, A.~M., {Barklem}, P.~S., {Asplund}, M., {Collet}, R., \&
  {Zatsarinny}, O. 2018{\natexlab{a}}, \aap, 616, A89

\bibitem[{{Amarsi} {et~al.}(2016{\natexlab{b}}){Amarsi}, {Lind}, {Asplund},
  {Barklem}, \& {Collet}}]{2016MNRAS.463.1518A}
{Amarsi}, A.~M., {Lind}, K., {Asplund}, M., {Barklem}, P.~S., \& {Collet}, R.
  2016{\natexlab{b}}, \mnras, 463, 1518

\bibitem[{{Amarsi} {et~al.}(2018{\natexlab{b}}){Amarsi}, {Nordlander},
  {Barklem}, {Asplund}, {Collet}, \& {Lind}}]{2018A&A...615A.139A}
{Amarsi}, A.~M., {Nordlander}, T., {Barklem}, P.~S., {et~al.}
  2018{\natexlab{b}}, \aap, 615, A139

\bibitem[{{Asplund} {et~al.}(2009){Asplund}, {Grevesse}, {Sauval}, \&
  {Scott}}]{2009ARA&amp;A..47..481A}
{Asplund}, M., {Grevesse}, N., {Sauval}, A.~J., \& {Scott}, P. 2009, \araa, 47,
  481

\bibitem[{{Barklem}(2016)}]{2016A&amp;ARv..24....9B}
{Barklem}, P.~S. 2016, \aapr, 24, 9

\bibitem[{{Casagrande} {et~al.}(2010){Casagrande}, {Ram{\'{\i}}rez},
  {Mel{\'e}ndez}, {Bessell}, \& {Asplund}}]{2010A&amp;A...512A..54C}
{Casagrande}, L., {Ram{\'{\i}}rez}, I., {Mel{\'e}ndez}, J., {Bessell}, M., \&
  {Asplund}, M. 2010, \aap, 512, A54

\bibitem[{{Cescutti} {et~al.}(2009){Cescutti}, {Matteucci}, {McWilliam}, \&
  {Chiappini}}]{2009A&amp;A...505..605C}
{Cescutti}, G., {Matteucci}, F., {McWilliam}, A., \& {Chiappini}, C. 2009,
  \aap, 505, 605

\bibitem[{{Chiappini} {et~al.}(2006){Chiappini}, {Hirschi}, {Meynet},
  {Ekstr{\"o}m}, {Maeder}, \& {Matteucci}}]{2006A&A...449L..27C}
{Chiappini}, C., {Hirschi}, R., {Meynet}, G., {et~al.} 2006, \aap, 449, L27

\bibitem[{{Chiappini} {et~al.}(2003){Chiappini}, {Romano}, \&
  {Matteucci}}]{2003MNRAS.339...63C}
{Chiappini}, C., {Romano}, D., \& {Matteucci}, F. 2003, \mnras, 339, 63

\bibitem[{{Collet} {et~al.}(2006){Collet}, {Asplund}, \&
  {Trampedach}}]{2006ApJ...644L.121C}
{Collet}, R., {Asplund}, M., \& {Trampedach}, R. 2006, \apjl, 644, L121

\bibitem[{{Collet} {et~al.}(2018){Collet}, {Nordlund}, {Asplund}, {Hayek}, \&
  {Trampedach}}]{2018MNRAS.475.3369C}
{Collet}, R., {Nordlund}, {\~A}., {Asplund}, M., {Hayek}, W., \& {Trampedach},
  R. 2018, \mnras, 475, 3369

\bibitem[{{Dobrovolskas} {et~al.}(2015){Dobrovolskas}, {Ku{\v c}inskas},
  {Bonifacio}, {Caffau}, {Ludwig}, {Steffen}, \&
  {Spite}}]{2015A&amp;A...576A.128D}
{Dobrovolskas}, V., {Ku{\v c}inskas}, A., {Bonifacio}, P., {et~al.} 2015, \aap,
  576, A128

\bibitem[{{Fabbian} {et~al.}(2009){Fabbian}, {Nissen}, {Asplund}, {Pettini}, \&
  {Akerman}}]{2009A&amp;A...500.1143F}
{Fabbian}, D., {Nissen}, P.~E., {Asplund}, M., {Pettini}, M., \& {Akerman}, C.
  2009, \aap, 500, 1143

\bibitem[{{Gaia Collaboration} {et~al.}(2018){Gaia Collaboration}, {Brown},
  {Vallenari}, {Prusti}, {de Bruijne}, {Babusiaux}, {Bailer-Jones}, {Biermann},
  {Evans}, {Eyer}, \& et~al.}]{2018A&A...616A...1G}
{Gaia Collaboration}, {Brown}, A.~G.~A., {Vallenari}, A., {et~al.} 2018, \aap,
  616, A1

\bibitem[{{Gallagher} {et~al.}(2017){Gallagher}, {Caffau}, {Bonifacio},
  {Ludwig}, {Steffen}, {Homeier}, \& {Plez}}]{2017A&amp;A...598L..10G}
{Gallagher}, A.~J., {Caffau}, E., {Bonifacio}, P., {et~al.} 2017, \aap, 598,
  L10

\bibitem[{{Grevesse} {et~al.}(2007){Grevesse}, {Asplund}, \&
  {Sauval}}]{2007coma.book..105G}
{Grevesse}, N., {Asplund}, M., \& {Sauval}, A.~J. 2007, {The Solar Chemical
  Composition}, ed. R.~{von Steiger}, G.~{Gloeckler}, \& G.~M. {Mason}
  (Springer Science+Business Media), 105

\bibitem[{{Gustafsson} {et~al.}(2008){Gustafsson}, {Edvardsson}, {Eriksson},
  {J{\o}rgensen}, {Nordlund}, \& {Plez}}]{2008A&amp;A...486..951G}
{Gustafsson}, B., {Edvardsson}, B., {Eriksson}, K., {et~al.} 2008, \aap, 486,
  951

\bibitem[{{Hibbert} {et~al.}(1991){Hibbert}, {Biemont}, {Godefroid}, \&
  {Vaeck}}]{1991JPhB...24.3943H}
{Hibbert}, A., {Biemont}, E., {Godefroid}, M., \& {Vaeck}, N. 1991, Journal of
  Physics B Atomic Molecular Physics, 24, 3943

\bibitem[{{Hibbert} {et~al.}(1993){Hibbert}, {Biemont}, {Godefroid}, \&
  {Vaeck}}]{1993A&AS...99..179H}
{Hibbert}, A., {Biemont}, E., {Godefroid}, M., \& {Vaeck}, N. 1993, \aaps, 99,
  179

\bibitem[{{Ishigaki} {et~al.}(2014){Ishigaki}, {Tominaga}, {Kobayashi}, \&
  {Nomoto}}]{2014ApJ...792L..32I}
{Ishigaki}, M.~N., {Tominaga}, N., {Kobayashi}, C., \& {Nomoto}, K. 2014,
  \apjl, 792, L32

\bibitem[{{Jacobson} \& {Frebel}(2015)}]{2015ApJ...808...53J}
{Jacobson}, H.~R. \& {Frebel}, A. 2015, \apj, 808, 53

\bibitem[{{Karakas} \& {Lattanzio}(2014)}]{2014PASA...31...30K}
{Karakas}, A.~I. \& {Lattanzio}, J.~C. 2014, \pasa, 31, e030

\bibitem[{{Kobayashi} \& {Nomoto}(2009)}]{2009ApJ...707.1466K}
{Kobayashi}, C. \& {Nomoto}, K. 2009, \apj, 707, 1466

\bibitem[{Kramida {et~al.}(2015)Kramida, {Yu.~Ralchenko}, Reader, \& {and NIST
  ASD Team}}]{NIST_ASD}
Kramida, A., {Yu.~Ralchenko}, Reader, J., \& {and NIST ASD Team}. 2015, NIST,
  {NIST Atomic Spectra Database (ver. 5.3), [Online]. Available:
  {\url{http://physics.nist.gov/asd}} [2015, November 2]. National Institute of
  Standards and Technology, Gaithersburg, MD.}

\bibitem[{{Leenaarts} \& {Carlsson}(2009)}]{2009ASPC..415...87L_short}
{Leenaarts}, J. \& {Carlsson}, M. 2009, in Astronomical Society of the Pacific
  Conference Series, Vol. 415, The Second Hinode Science Meeting, ed.
  B.~{Lites}, M.~{Cheung}, T.~{Magara}, J.~{Mariska}, \& K.~{Reeves}, 87

\bibitem[{{Magic} {et~al.}(2013){Magic}, {Collet}, {Asplund}, {Trampedach},
  {Hayek}, {Chiavassa}, {Stein}, \& {Nordlund}}]{2013A&amp;A...557A..26M}
{Magic}, Z., {Collet}, R., {Asplund}, M., {et~al.} 2013, \aap, 557, A26

\bibitem[{{Mel{\'e}ndez} \& {Barbuy}(2009)}]{2009A&amp;A...497..611M}
{Mel{\'e}ndez}, J. \& {Barbuy}, B. 2009, \aap, 497, 611

\bibitem[{{Meynet} {et~al.}(2006){Meynet}, {Ekstr{\"o}m}, \&
  {Maeder}}]{2006A&A...447..623M}
{Meynet}, G., {Ekstr{\"o}m}, S., \& {Maeder}, A. 2006, \aap, 447, 623

\bibitem[{{Nissen} {et~al.}(2007){Nissen}, {Akerman}, {Asplund}, {Fabbian},
  {Kerber}, {Kaufl}, \& {Pettini}}]{2007A&amp;A...469..319N}
{Nissen}, P.~E., {Akerman}, C., {Asplund}, M., {et~al.} 2007, \aap, 469, 319

\bibitem[{{Nissen} {et~al.}(2004){Nissen}, {Chen}, {Asplund}, \&
  {Pettini}}]{2004A&amp;A...415..993N}
{Nissen}, P.~E., {Chen}, Y.~Q., {Asplund}, M., \& {Pettini}, M. 2004, \aap,
  415, 993

\bibitem[{{Nissen} {et~al.}(2014){Nissen}, {Chen}, {Carigi}, {Schuster}, \&
  {Zhao}}]{2014A&amp;A...568A..25N}
{Nissen}, P.~E., {Chen}, Y.~Q., {Carigi}, L., {Schuster}, W.~J., \& {Zhao}, G.
  2014, \aap, 568, A25

\bibitem[{{Nissen} {et~al.}(2002){Nissen}, {Primas}, {Asplund}, \&
  {Lambert}}]{2002A&amp;A...390..235N}
{Nissen}, P.~E., {Primas}, F., {Asplund}, M., \& {Lambert}, D.~L. 2002, \aap,
  390, 235

\bibitem[{{Nissen} \& {Schuster}(2010)}]{2010A&amp;A...511L..10N}
{Nissen}, P.~E. \& {Schuster}, W.~J. 2010, \aap, 511, L10

\bibitem[{{Pinsonneault}(1997)}]{1997ARA&amp;A..35..557P}
{Pinsonneault}, M. 1997, \araa, 35, 557

\bibitem[{{Placco} {et~al.}(2016){Placco}, {Beers}, {Reggiani}, \&
  {Mel{\'e}ndez}}]{2016ApJ...829L..24P}
{Placco}, V.~M., {Beers}, T.~C., {Reggiani}, H., \& {Mel{\'e}ndez}, J. 2016,
  \apjl, 829, L24

\bibitem[{{Ram{\'{\i}}rez} {et~al.}(2012){Ram{\'{\i}}rez}, {Mel{\'e}ndez}, \&
  {Chanam{\'e}}}]{2012ApJ...757..164R}
{Ram{\'{\i}}rez}, I., {Mel{\'e}ndez}, J., \& {Chanam{\'e}}, J. 2012, \apj, 757,
  164

\bibitem[{{Reggiani} {et~al.}(2017){Reggiani}, {Mel{\'e}ndez}, {Kobayashi},
  {Karakas}, \& {Placco}}]{2017A&amp;A...608A..46R}
{Reggiani}, H., {Mel{\'e}ndez}, J., {Kobayashi}, C., {Karakas}, A., \&
  {Placco}, V. 2017, \aap, 608, A46

\bibitem[{{Roederer} {et~al.}(2014){Roederer}, {Preston}, {Thompson},
  {Shectman}, \& {Sneden}}]{2014ApJ...784..158R}
{Roederer}, I.~U., {Preston}, G.~W., {Thompson}, I.~B., {Shectman}, S.~A., \&
  {Sneden}, C. 2014, \apj, 784, 158

\bibitem[{{Smiljanic} {et~al.}(2008){Smiljanic}, {Pasquini}, {Primas},
  {Mazzali}, {Galli}, \& {Valle}}]{2008MNRAS.385L..93S}
{Smiljanic}, R., {Pasquini}, L., {Primas}, F., {et~al.} 2008, \mnras, 385, L93

\bibitem[{{Tinsley}(1979)}]{1979ApJ...229.1046T}
{Tinsley}, B.~M. 1979, \apj, 229, 1046

\bibitem[{{VandenBerg} {et~al.}(2014){VandenBerg}, {Bond}, {Nelan}, {Nissen},
  {Schaefer}, \& {Harmer}}]{2014ApJ...792..110V}
{VandenBerg}, D.~A., {Bond}, H.~E., {Nelan}, E.~P., {et~al.} 2014, \apj, 792,
  110

\end{thebibliography}


\label{lastpage}
\end{document}